\documentclass[10pt,conference]{IEEEtran}
\usepackage{braket}
\usepackage{amsmath}
\usepackage{graphicx}
\usepackage{cite}
\usepackage{mathtools}
\usepackage{amsfonts}
\usepackage{url}
\usepackage{bbm}
\usepackage{float}
\usepackage{stfloats}
\usepackage{tabularx}
\usepackage{xcolor}

\everypar{\looseness=-1}

\newcommand{\tr}{\text{Tr}}
\newcommand{\ketbra}[1]{\ket{#1}\bra{#1}}
\newcommand{\bigchi}{\makebox{\large\ensuremath{\chi}}}

\title{A Three-Mode Erasure Code for Continuous Variable Quantum Communications}

\author{
	 \IEEEauthorblockN{Eduardo Villase\~nor and Robert Malaney}\\
	 \IEEEauthorblockA{School of Electrical Engineering  \& Telecommunications,\\
		 The University of New South Wales, Sydney, NSW 2052, Australia.\\
     }
}
\IEEEoverridecommandlockouts\IEEEpubid{\makebox[\columnwidth]{
\begin{tabular}[t]{@{}l@{}}This work has been acceped for publication in GLOBECOM 2022.\\Copyright: 978-1-6654-3540-6/22~\copyright~2022 IEEE \end{tabular} \hfill} \hspace{\columnsep}\makebox[\columnwidth]{ }}
\begin{document}

\maketitle

\begin{abstract}
Quantum states of light being transmitted via realistic free-space channels often suffer erasure errors due to several factors such as coupling inefficiencies between transmitter and receiver. In this work, an error correction code capable of protecting a single-mode quantum state against erasures is presented. Our three-mode code protects a single-mode Continuous Variable (CV) state via a bipartite CV entangled state. 
In realistic deployments, it can almost completely reverse a single erasure on the encoded state, and for two erasures can it improve the fidelities of received states relative to direct transmission. The bipartite entangled state used in the encoding can be Gaussian or non-Gaussian, with the latter further enhancing the performance of the code.
Our new code is the simplest code known that protects a single mode against erasures and should prove useful in the construction of practical CV quantum networks that rely on free-space optics. 
\end{abstract}
\begin{IEEEkeywords}
Continuous variable quantum information, quantum communications, quantum error correction.
\end{IEEEkeywords}
%\vspace{-3cm}
\IEEEpeerreviewmaketitle

\section{Introduction}

Continuous variable (CV) quantum information, encoded in the quadrature variables of electromagnetic signals, may offer several advantages over discrete variable (DV) quantum information in the context of reliable-state transfer over free space \cite{GaussianQuantumInformation, doi:10.1142/S1230161214400010}. The free-space transmission of CV quantum states using weak laser pulses via satellites in low-Earth-orbit (LEO), could potentially represent a viable path toward achieving global quantum communications \cite{PhysRevA.100.012325,PhysRevA.95.022312, NedaReview}.

However, Quantum information is fragile by nature - the unavoidable interaction between quantum systems and their environment introduces errors to the quantum states. Correction of these errors requires the use of additional quantum resources to construct a larger quantum system in which the deterministic identification and correction of the errors is possible \cite{Fukui_2022}. Attempts at error correction in many contexts for CV states have been attempted, e.g. \cite{PhysRevA.64.012310, PhysRevX.10.011058,nonGaussianTeleportation,DellAnno1,MingjianPAPS, ScissorsQKD, 9024548, PhysRevA.91.063832, 9463774}.
Here we focus on a type of error that can affect CV states in practical scenarios, especially with the advent of new communications platforms such as quantum communications via satellite; erasure channels.

In the context of free-space quantum communications, the erasure channel corresponds to the beam being completely lost during transmission. This can happen as a consequence of the beam wandering effects caused by the turbulence in the atmosphere \cite{PhysRevLett.117.090501, 9348086, Sebastian}. The uplink transmission of quantum states from the ground to a LEO satellite is an example where erasures are prominent due to the prevailing beam wandering \cite{9463774}.

To achieve the correction of erasures on quantum states a
new quantum erasure code is presented.  The code considers a single-mode quantum state as an input and encodes it with a bipartite entangled state.
Our code is different from erasure codes previously constructed. The most similar code to that presented here is the code of \cite{PhysRevLett.101.130503, Lassen2010} in which two input states are protected through the use of four transmission channels. In contrast, our code protects fewer states (one) but with the benefit of reduced complexity (use of three transmission channels). As such, our code offers a pathway to more pragmatic deployments.  
In addition, due to its relative simplicity,  it becomes possible to optimize faster the free parameters of our code relative to other codes - an issue of particular importance when multiple erasures occur on the encoded state.
The novel contributions of this work are summarized as follows:
\begin{itemize}
  \item We present an erasure code for CV states and analyze in detail its performance.
 \item  A detailed optimization procedure is presented for our code when erasures are present. Additionally, performance with a simpler deployment where this optimization is neglected is compared.
 \item  The use of  Gaussian and non-Gaussian states for the input entangled state is considered and their performance is compared. It is shown that, in combination with the optimization process, the use of non-Gaussian states further increases the performance of the code.
\end{itemize}

In section~II we introduce our code and analyze its performance via the Wigner Characteristic Function (CF) formalism. We present detailed results from the code which detail its performance under different combinations of erasures and different assumptions regarding the input entangled state used for encoding in section~III. Finally, we draw our conclusions in section~IV.

\section{Erasure error correction code}
The Wigner CF formalism will be used to study the error correction code presented here.
This is motivated by the results presented in \cite{MarianMarian} that show that the output state of CV quantum teleportation can be easily computed from the CFs of the quantum states involved in the protocol. In this work a similar result is obtained, the CF of the output state after error correction corresponds to a product of the CFs of the input state and the entangled state.

The CF of any $n$-mode quantum state $\hat{\rho}$ is defined as
\begin{align}
\bigchi(\lambda_1, \lambda_2, ..., \lambda_n) = \tr\left\{\hat{\rho} \hat{D}(\lambda_1)\hat{D}(\lambda_2)...\hat{D}(\lambda_n)\right\},
\end{align}
where $\lambda_i \in \mathbb{C}$. Here, $\hat{D}$ is the displacement operator,
\begin{align}
\hat{D}(\lambda_i) = e^{\lambda_j \hat{a}^{\dag}_j - \lambda_j^* \hat{a}_j},
\label{eq:displacement}
\end{align}
where $\hat{a}_j$ and $\hat{a}^{\dag}_j$ are the annihilation and creation operators of mode $j$, and ${}^*$ represents the complex conjugate.
Conveniently, linear optics operations can be expressed in the CF formalism by simply transforming the arguments of the CF, whilst leaving the functions
themselves unchanged.
Relevant CFs for this work include  the vacuum state, $\ket{0}$, expressed as
\begin{align}
\bigchi_{\ket{0}}(\lambda) = \exp \left[ -\frac{|\lambda|^2}{2} \right],
\end{align}
and  the coherent state, $\ket{\alpha} = \hat{D}(\alpha)\ket{0}$,   expressed as
\begin{align}
  \bigchi_{\ket{\alpha}}(\lambda) = \exp \left[-\frac{|\lambda|^2}{2} + (\lambda \alpha^* - \lambda^* \alpha) \right].
\end{align}
The other important CFs we will utilize involve those of the different entangled states used in the encoding, which we present later.

\begin{figure*}
  \centering
  \includegraphics[width=.95\textwidth ]{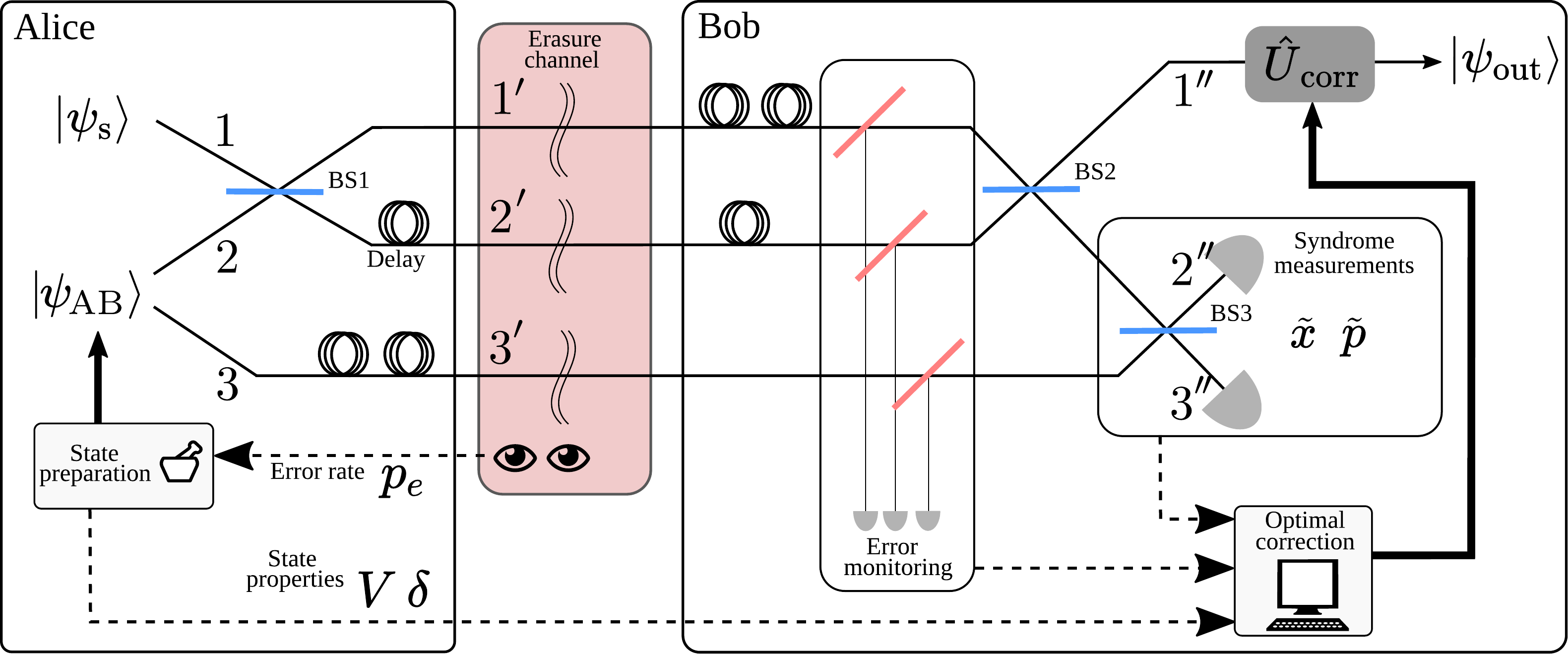}
  \caption{The code to correct the erasures introduced by the quantum erasure channel. The code protects a single-mode quantum state against erasures by combining it with an entangled bipartite state. The encoding and decoding are made using beam splitters. Delay lines are introduced such that each mode is sent independently through the channel. The syndrome measurement corresponds to a dual homodyne measurement. Using the syndrome result a correction is applied to the remaining mode of the state to recover the original state. The error code is optimized by monitoring another (classical) beam in the same channels  (red BSs) and preparing the entangled state accordingly. Dashed lines represent the transmission of classical information.}
  \label{fig:qec}
\end{figure*}

In Fig.~\ref{fig:qec} our error correction code  is presented. The resources required for the implementation and optimization of this code include entangled-bipartite states, linear-optics operations,  and classical processing.
The deployment of our error correction scheme can be divided into four steps: encoding, decoding, syndrome measurements, and correction.

As shown in Fig.~\ref{fig:qec}, Alice starts with a single-mode quantum state, which we will refer to as the ``quantum signal,'' that she wishes to transmit through the channel to Bob, and prepares an entangled bipartite state. The initial CF corresponds to a product of the CFs of the quantum signal, $\bigchi_\mathrm{s}(\lambda_1)$, and the entangled state, $\bigchi_\mathrm{AB}(\lambda_2, \lambda_3)$.
The encoding of the quantum signal is done via a balanced beam splitter (BS1), described by a transformation of the CF arguments as follows \cite{DellAnno1},
\begin{align}
  \bigchi_\mathrm{s}\left( \frac{\lambda_1 + \lambda_2}{\sqrt{2}} \right) \bigchi_\mathrm{AB} \left(\frac{\lambda_1 - \lambda_2}{\sqrt{2}}, \lambda_3 \right).
  \label{eq:bs}
\end{align}

Thereafter, the encoded state is transmitted from Alice to Bob through the channel.
In general, the erasure channel acting on a single-mode state, $\rho$, returns the state,
\begin{align}
  \rho' = (1-P_e) \rho + P_e\ketbra{0},
\end{align}
with $P_e$ being the probability of an erasure.
If the three modes of the encoded state were sent concomitantly through the channel the result would be either an unchanged state or a three-mode vacuum state from which no information can be recovered. Therefore, a mechanism that transmits each mode independently must be used. An example of such a mechanism would be one that time multiplexes each mode using delay lines.
When the three modes are sent independently through the channel the result is a mixed state corresponding to all of the eight combinations of modes erased,
\begin{align}
  \rho_\mathrm{ch} = \sum_{j=1}^8 P_j \rho_j,
\end{align}
with $P_j$ corresponding to the probability of each combination of modes erased. These range from $(1-P_e)^3$ to $P_e^3$, corresponding to zero erasures, and erasures in every mode, respectively.
At this point, the CF of the three mode state, now defined as $\bigchi_\mathrm{ch}^\mathbf{x}(\lambda_1, \lambda_2, \lambda_3)$, will depend on the modes $\mathbf{x}$ that suffer an erasure. This means that for the erased modes their arguments in the CF in Eq.~\ref{eq:bs} will be set to zero \cite{https://doi.org/10.48550/arxiv.quant-ph/0612029}, while vacuum CFs are added as a product with arguments corresponding to the erased modes. For example, if mode $2'$ has an erasure the corresponding CF is
\begin{align}
  \bigchi_\mathrm{ch}^\mathrm{\{2\}}(\lambda_1, \lambda_2, \lambda_3) = \bigchi_\mathrm{s}\left(\frac{\lambda_1}{\sqrt{2}} \right) \bigchi_\mathrm{AB}\left(\frac{\lambda_1}{\sqrt{2}},  \lambda_3\right) \bigchi_{\ket{0}}(\lambda_2).
\end{align}
Bob monitors and identifies which of the modes suffered an erasure during transmission via the channel, and will use that information in combination with the syndrome measurement to apply the correction. Note, our erasure code implementation resembles that utilized in the context of secret sharing \cite{sharing}. Apart from the application context, a key difference in our implementation is the use of the erasure monitoring function and its mapping to an erasure correction protocol. We can consider this mapping to be the following: for any monitoring measurement that indicates a non-unity channel transmissivity, we set that channel to be in `erasure'. This logic is then used to set the gains needed to adjust the output quantum state.

As the modes are received by Bob, he applies corresponding delays such that by the time he has received all the modes, they are all temporally coincident.
To decode the quantum state, he applies BS2 that transforms the arguments as,
\begin{align}
  \bigchi_\mathrm{ch}^\mathbf{x}\left( \frac{\lambda_1 + \lambda_2}{\sqrt{2}}, \frac{\lambda_1 - \lambda_2}{\sqrt{2}}, \lambda_3 \right).
\end{align}
In the case when the channel acts as an identity (no erasures) the application of BS2 effectively cancels the effects of BS1.

During the final step, syndrome measurements are performed. To this end, BS3 is applied, giving a CF for the three mode state as
\begin{align}
\label{eq:cfbs}
  \bigchi_\mathrm{BS3}^\mathbf{x}&(\lambda_1, \lambda_2, \lambda_3)= \\
  &\bigchi_\mathrm{ch}^\mathbf{x}\left( \frac{\lambda_1 + \lambda_2}{\sqrt{2}}, \frac{\lambda_1}{\sqrt{2}}- \frac{\lambda_2 + \lambda_3}{2}, \frac{\lambda_2 - \lambda_3}{\sqrt{2}}\right).   \nonumber
  %\\ \nonumber
\end{align}
Now dual homodyne measurements are performed on modes $2''$ and $3''$ (see Fig.~\ref{fig:qec}).
To compute the result after the measurements it is convenient to represent the complex arguments in the {\it phase-space} representation,  by using two distinct real numbers, $x_j {=} \frac{1}{\sqrt{2}}(\lambda_j + \lambda_j^*)$ and $p_j{=} \frac{i}{\sqrt{2}}(\lambda_j^* - \lambda_j)$.
Then, for the pair of measurement results $\tilde{x}$ and $\tilde{p}$ the output CF (on mode $1''$) is obtained by the integration \cite{MarianMarian},
\begin{align}
  \bigchi_\mathrm{m}^\mathbf{x}(x, p) &= \frac{\mathcal{P}(\tilde{x}, \tilde{p})^{-1}}{(2\pi)^2} \int d x_2 d p_3 \bigchi_\mathrm{BS3}^\mathbf{x}\left(x, p, x_2, 0, 0, p_3 \right) \nonumber \\ &~~~~~~~~~~~~~~~~~~~\times e^{-i\tilde{x}p_3 + i\tilde{p}x_2},
\label{eq:CFm}
\end{align}
with $\mathcal{P}(\tilde{x}, \tilde{p})$ the probability distribution of any pair of measurement results, given by
\begin{align}
  \mathcal{P}(\tilde{x}, \tilde{p} )= \frac{1}{(2\pi)^2} \int d x_2 d p_3 \bigchi_\mathrm{BS3}^\mathbf{x}\left(0, 0, x_2, 0, 0, p_3 \right) \nonumber \\
  \times e^{-i\tilde{x}p_3 + i\tilde{p}x_2}.
\end{align}
The extra exponential term in Eq.~\ref{eq:CFm} indicates that the collapsed state after the measurement requires a corrective displacement to be recovered. The corrective displacement is,
\begin{align}
\hat{U}_\mathrm{corr}(\tilde{x}, \tilde{p}) = \exp\left[\sqrt{2}\tilde{x}p g_p - i\sqrt{2}\tilde{p}x g_x\right],
\end{align}
where $(g_x,g_p)$ are the gains of the correction. Here the correction includes a factor of $\sqrt{2}$ to compensate for the global factor that appears in the arguments of $\bigchi_\mathrm{ch}^\mathbf{x}$ in Eq.~\ref{eq:cfbs}.

Finally, since the output CF will depend on a specific set of measurement results,
the mean output over all possible measurement outcomes weighted with the corresponding probability distribution must be considered, that is
\begin{align}
\bigchi_\mathrm{out}^\mathbf{x}(x, p) &= \int d \tilde{x} d \tilde{p}
\mathcal{P}(\tilde{x}, \tilde{p}) \bigchi_\mathrm{m}^\mathbf{x}(x, p) e^{i\sqrt{2}\tilde{x}p g_p - i\sqrt{2} \tilde{p} x g_x} \nonumber \\
&= \frac{1}{(2 \pi)^2} \int d \tilde{x} d \tilde{p} d x_2 d p_3
 \bigchi_\mathrm{BS3}^\mathbf{x} (x, p,x_2, 0, 0, p_3 ) \nonumber \\ & ~~~~~~~~~~\times e^{i \tilde{x} (\sqrt{2} p g_p - p_3) - i \tilde{p} ( \sqrt{2} x g_x - x_2)}.
\end{align}
At this point the definition of the Dirac delta function, $\frac{1}{2\pi}\iint e^{i\beta x -i\beta \alpha}f(\alpha)d\beta d\alpha = \int \delta (x - \alpha)f(\alpha)d\alpha$, can be used twice to finally obtain
\begin{align}
  \bigchi_\mathrm{out}^\mathbf{x}(x, p)&= \int d x_2 d p_3 \delta (p_3 - \sqrt{2}g_p p)\delta (x_2 -\sqrt{2}g_x x) \nonumber \\
  & ~~~~~~~~\times \bigchi_\mathrm{BS3}^\mathbf{x}(x, p, x_2, 0, 0, p_3 ).
\end{align}
Therefore, the resulting CF will correspond to a product of the initial CFs with a transformation of the parameters that will depend on the specific erasures $\mathbf{x}$ plus a vacuum contribution.
 The resulting expressions for $\bigchi_\mathrm{out}^\mathbf{x}$ are summarized in Table~I.
The CF of the output state depends on the erasures, which can be divided into two cases:
\subsubsection{Single erasure}
If there is only a single erasure on the encoded state, the signal is recovered by maximizing the amount of entanglement in $\ket{\rho_\mathrm{AB}}$, with perfect recovery corresponding to the limit of infinite entanglement. To this end, see that under the correct choice of $(g_x, g_p)$ (e.g. if erasure on mode $2'$, then $g_x{=}g_p{=}1$) the CF is reduced to a product of the CFs of the quantum signal and the entangled state. The signal is recovered since $V{\rightarrow}\infty$ implies that $\bigchi_\mathrm{AB}(\lambda, \lambda^*) {\rightarrow} 1$. When the erasure is on mode $3'$, the signal can always be recovered fully independently of the entanglement used by setting $g_x{=}g_p{=}0$.

%Add subsections
\subsubsection{Two erasures}
When two modes suffer an erasure, any amount of entanglement in $\ket{\rho_\mathrm{AB}}$ will ultimately become added noise in the output state. This can be seen from the fact that the CF of the entangled state $\bigchi_\mathrm{AB}$ has one mode traced out, meaning that the entangled state has now been reduced to a thermal state. Remarkably, some information from the signal can still be recovered in this case if the entangled state is instead replaced by two vacuum states \cite{PhysRevLett.101.130503}.

Ultimately, in a realistic scenario, there might be a single or two erasures. In such a scenario there is an optimal amount of entanglement in the state $\ket{\rho_\mathrm{AB}}$, that depends on $P_e$, that will balance the trade-off between recovering the signal when there is a single erasure, and reducing the noise when there are two erasures.

\begin{table*}[b]
\centerline{{\bf Table I - Output state CF}}
\begin{center}
\begin{tabularx}{0.8\textwidth}{lc}
 Erasure on mode & $\bigchi_\mathrm{out}^\mathbf{x}$\\  \noalign{\vskip 0.5mm}
 \hline  \noalign{\vskip 0.5mm}
 none & $\bigchi_\mathrm{s}(x,p)$ \\ [.2cm]
 $1'$ & $\bigchi_\mathrm{s}\left(\frac{1-g_x}{2} x, \frac{1-g_p}{2} p\right) \bigchi_\mathrm{AB}\left(-\frac{1-g_x}{2} x, -\frac{1-g_p}{2} p, g_x x, -g_p p\right) \bigchi_{\ket{0}}\left(\frac{1+g_x}{\sqrt{2}} x, \frac{1+g_p}{\sqrt{2}} p\right)$ \\ [.2cm]
 $2'$ & $\bigchi_\mathrm{s}\left(\frac{1+g_x}{2} x, \frac{1+g_p}{2} p\right) \bigchi_\mathrm{AB}\left(\frac{1+g_x}{2} x, \frac{1+g_p}{2} p, g_x x, -g_p p\right) \bigchi_{\ket{0}}\left(\frac{1-g_x}{\sqrt{2}} x, \frac{1-g_p}{\sqrt{2}} p\right)$ \\ [.2cm]
 $3'$ & $\bigchi_\mathrm{s}(x, p) \bigchi_\mathrm{AB}\left(g_x x, g_p p, 0, 0\right) \bigchi_{\ket{0}}\left(g_x x, -g_p p\right)$ \\ [.2cm]
 $1'$ \& $2'$ & $\bigchi_\mathrm{AB}\left(0,0, g_x x, -g_p p\right) \bigchi_{\ket{0}}\left(\frac{1+g_x}{\sqrt{2}} x, \frac{1+g_p}{\sqrt{2}} p\right) \bigchi_{\ket{0}}\left(\frac{1-g_x}{\sqrt{2}} x, \frac{1-g_p}{\sqrt{2}} p\right)$ \\ [.2cm]
 $1'$ \& $3'$ & $\bigchi_\mathrm{s}\left(\frac{1-g_x}{2} x, \frac{1-g_p}{2} p\right) \bigchi_\mathrm{AB}\left(-\frac{1-g_x}{2} x, -\frac{1-g_p}{2} p, 0, 0\right) \bigchi_{\ket{0}}\left(\frac{1+g_x}{\sqrt{2}} x, \frac{1+g_p}{\sqrt{2}} p\right) \bigchi_{\ket{0}}\left(\frac{1+g_x}{\sqrt{2}} x, \frac{1-g_p}{\sqrt{2}} p\right)$ \\ [.2cm]
 $2'$ \& $3'$ & $\bigchi_\mathrm{s}\left(\frac{1+g_x}{2} x, \frac{1+g_p}{2} p\right) \bigchi_\mathrm{AB}\left(\frac{1+g_x}{2} x, \frac{1+g_p}{2} p, 0, 0\right) \bigchi_{\ket{0}}\left(\frac{1-g_x}{\sqrt{2}} x, \frac{1-g_p}{\sqrt{2}} p\right) \bigchi_{\ket{0}}\left(g_x x, -g_p p\right)$ \\ [.2cm]
  $1'$ \& $2'$ \& $3'$ & $\bigchi_{\ket{0}}(x, p)$ \\
\end{tabularx}
\end{center}
\label{table:CF}
\end{table*}

\subsection{Optimizing the correction}
As discussed, the correction $\hat{U}_\mathrm{corr}$ and the characteristics of the entangled state used during encoding can be optimized to enhance the effectiveness of the code.
The process of optimizing the error correction code involves two distinct steps. One step is optimizing the entangled state, while the other is optimizing the parameters in the correction, $(g_x, g_p)$, for each of the mode-erasure combinations.
Note that the optimal values $(g_x, g_p)$, do not depend on the erasure probability $P_e$, however, they do depend on the entangled state. On the other hand, the optimal properties of the entangled state do depend on $P_e$. Thus, full optimization of the code requires a nested optimization problem. This nested optimization problem can be avoided if Alice and Bob prepare beforehand a dictionary with the optimal values $(g_x, g_p)$ as a function of the properties of the entangled state. In this case, the optimization process is as follows. First, Alice monitors the erasure channel to determine the value of $P_e$. Using this value she then prepares an optimal entangled state. Alice shares the properties of the entangled state with Bob via a classical channel. After Bob has received the three modes of the code he identifies the erasures on the encoded state and performs syndrome measurements. With these three pieces of information (the entangled state properties, the erasures, and the syndrome), he then uses the dictionary to obtain the optimal values of $(g_x,g_p)$ and utilizes them in the correction.

\subsection{Entangled resources}
In CV quantum protocols, the two-mode squeezed vacuum state (TMSV) represents the most accessible bipartite entangled state. States such as the TMSV state are commonly referred to as Gaussian since they can be described fully by the first two statistical moments of the quadratures \cite{GaussianQuantumInformation}.
The CF of a TMSV state can be conveniently written using the following Bogoliuvov transformation \cite{DellAnno1}:
\begin{align}
\label{eq:Bogo}
&\hat{S}_{12}(\varrho) \hat{D}(\lambda_1) \hat{D}(\lambda_2) \hat{S}^\dag_{12}(\varrho) = \hat{D}(\lambda_1') \hat{D}(\lambda_2'),  \\
&\lambda_j' = \cosh(r)\lambda_j + e^{i\phi} \sinh(r) \lambda_k^* ~~~~ j,k = 1,2; ~~ j\neq k. \nonumber
\end{align}
where $\hat{S}_{12}$ is the two-mode squeezing operator with $\varrho{=}re^{i \phi}$, throughout this work, the value $\phi {=} \pi$ is set. 
% Here (and throughout this work), we assume $\hbar{=}2$.
 The squeezing magnitude $r$ is directly proportional to the entanglement of the TMSV state, which can be quantified by the variance $V$, which relates to the squeezing as $V {=} \cosh(2r)$.
Using the transformation in Eq.~\ref{eq:Bogo} the CF of a TMSV can be written as,
\begin{align}
\bigchi_\mathrm{TMSV}(\lambda_{\mathrm{A}}, \lambda_{\mathrm{B}}) = \exp \left[ -\frac{1}{2}\Big(|\lambda_\mathrm{A}'|^2 + |\lambda_\mathrm{B}'|^2 \Big) \right].
\end{align}

In contrast to Gaussian states, non-Gaussian states cannot be fully described using statistical moments, instead, they require a full description in the form of a density operator or a Wigner function. In this work, a non-Gaussian entangled state that is obtained by the application of the squeezing operator to a Bell state is considered.
The normalized CF of a {\it squeezed Bell} (SB) state is \cite{DellAnno1},
\begin{align}
&\bigchi_\mathrm{SB}(\lambda_\mathrm{A}, \lambda_\mathrm{B}) = (\cos^2(\delta) + \sin^2(\delta))^{-1/2}
\nonumber \\
&\times \exp \left[ -\frac{1}{2}\left(|\lambda_\mathrm{A}'|^2 + |\lambda_\mathrm{B}'|^2 \right) \right]
  \Big[ \cos^2(\delta)  + 2 \cos(\delta)\sin(\delta)  \nonumber \\
  &~~~~~~~~\times \Re\{ \lambda_\mathrm{A}' \lambda_\mathrm{B}'   \} +  \sin^2(\delta) (1 - |\lambda_\mathrm{A}'|^2) (1 - |\lambda_\mathrm{B}'|^2)  \Big],
 \label{eq:sb}
\end{align}
where $\Re\{ z\}$ is the real part of $z$, and Eq.~\ref{eq:Bogo} is used.
The optimization of the properties of the entangled state, mentioned in Section II.A, is done in terms of the amount of entanglement of the states, that is, on the variance $V$ for both TMSV and SB states. Additionally, in the SB state, the parameter $\delta$ is also optimized.

\subsection{Fidelity of error correction}
To quantify the effectiveness of the error correction the fidelity of transmitted coherent states is used. The fidelity corresponds to a measurement of the closeness between the initial coherent states and the output states.
In the CF formalism this is computed as,
\begin{align}
\mathcal{F}(\alpha) = \frac{1}{\pi} \int d^2 \lambda \bigchi_\mathrm{\ket{\alpha}}(\lambda) \bigchi_\mathrm{out}(-\lambda).
\label{eq:fidelity}
\end{align}

Note that in this case, the fidelity will be dependent on the value of $\alpha$. Therefore, to accurately assess the effectiveness of the error correction code, a mean fidelity over an ensemble of coherent states must be considered. The ensemble is specified by the following distribution,
\begin{align}\label{eq:probcoherent}
P(\alpha) = \frac{1}{\sigma \pi} \exp\left[-\frac{|\alpha|^2}{\sigma}\right],
\end{align}
with $\sigma$ the variance of the distribution.
Therefore, the mean fidelity over the ensemble will be used, defined as
\begin{align}\label{eq:avefidelity}
\bar{\mathcal{F}}  = \int d{\alpha}^2 P(\alpha) \mathcal{F}(\alpha).
\end{align}

\section{Results}
The fidelities are computed using Table I, in combination with Eq.~\ref{eq:fidelity}. The mean fidelities over the distribution of coherent states are computed using Eq.~\ref{eq:avefidelity}, where the value $\sigma{=}10$ is fixed throughout this work. In the calculations presented here, the value of $V$ has the upper limit of $V_\mathrm{max}{=}9$.

\begin{figure}
  \centering
  \includegraphics[width=.48\textwidth]{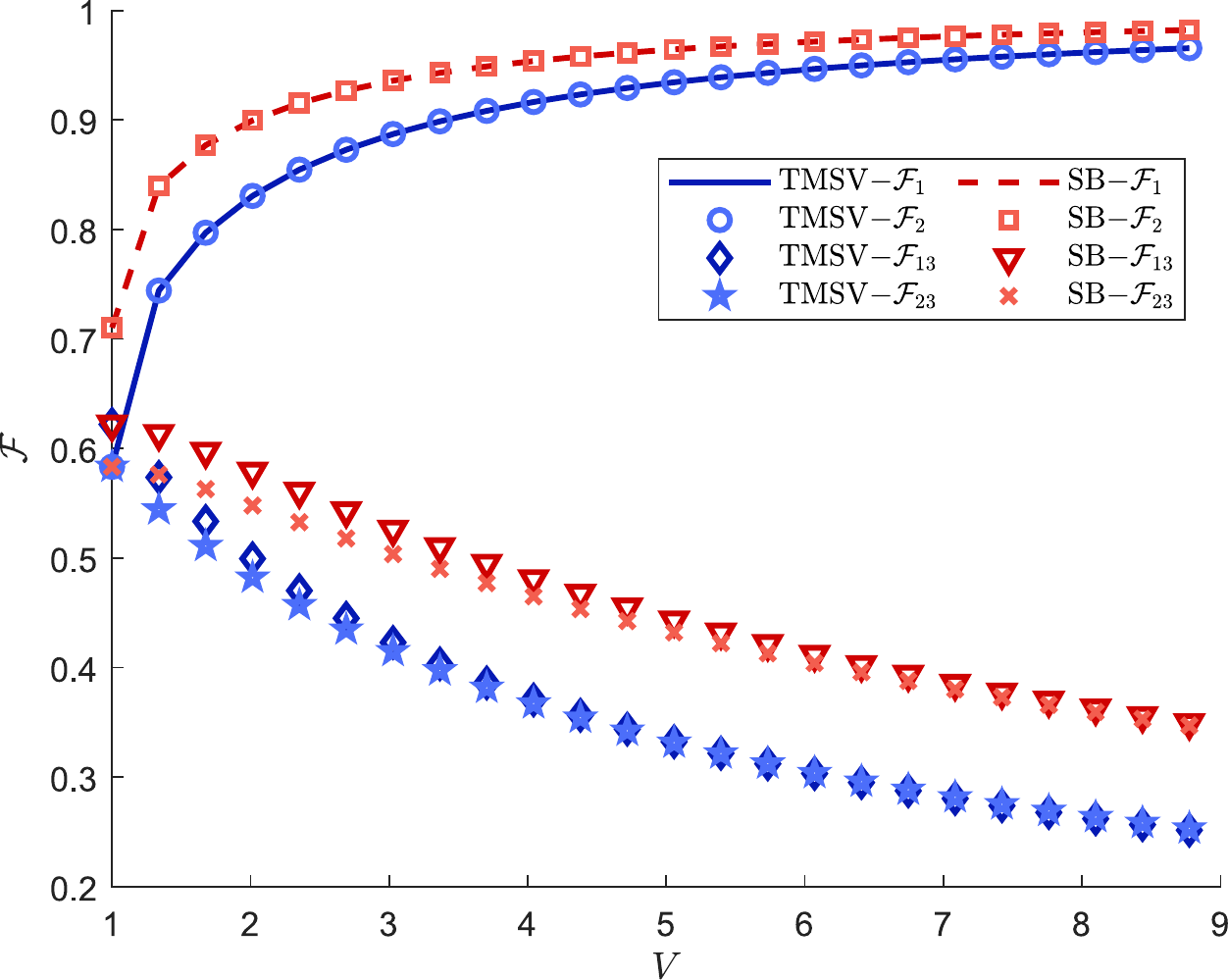}
  \caption{Fidelities corresponding to each possible combination of modes erased on the encoded state during transmission (see Table I). The fidelities are computed using a TMSV and a SB quantum state in the encoding. The optimal correction parameter $g_\mathrm{opt}$ is used in the calculations of the fidelities.
  }
  \label{fig:comp}
\end{figure}

To understand the output of the code, the fidelities corresponding to each case of the mode-erasure combinations are shown in Fig.~\ref{fig:comp}. The fidelities are presented as a function of $V$ of the entangled states, while for the SB state, the parameter $\delta$ is optimized for each value of $V$. In each case, the parameters $(g_x, g_p)$ are optimized. While in principle for any given coherent state the optimal values of $(g_x,g_p)$ are not equal, we find that when the optimization is done over the distribution of states given by Eq.~\ref{eq:probcoherent} the optimal values of both parameters coincide, that is $g_x {=} g_p {=}g_\mathrm{opt}$. See that the fidelities for the cases when modes $1'$ or $2'$ suffer an erasure tend to unity as $V$ increases. The particular case when mode $3'$ is lost is not shown in the figure, since the fidelity is always 1 when $g_\mathrm{opt} {=} 0$, as discussed above. 

Remarkably, in the cases where two modes are erased a higher fidelity than the classical limit of 0.5 can still be obtained if the entangled state is replaced by a vacuum, as discussed above. The combinations of modes erased that reduce the output to the vacuum state are omitted (simultaneous erasures in modes $1'$ and $2'$, and simultaneous erasures in all the modes).

\begin{figure}
  \centering
  \includegraphics[width=.48\textwidth]{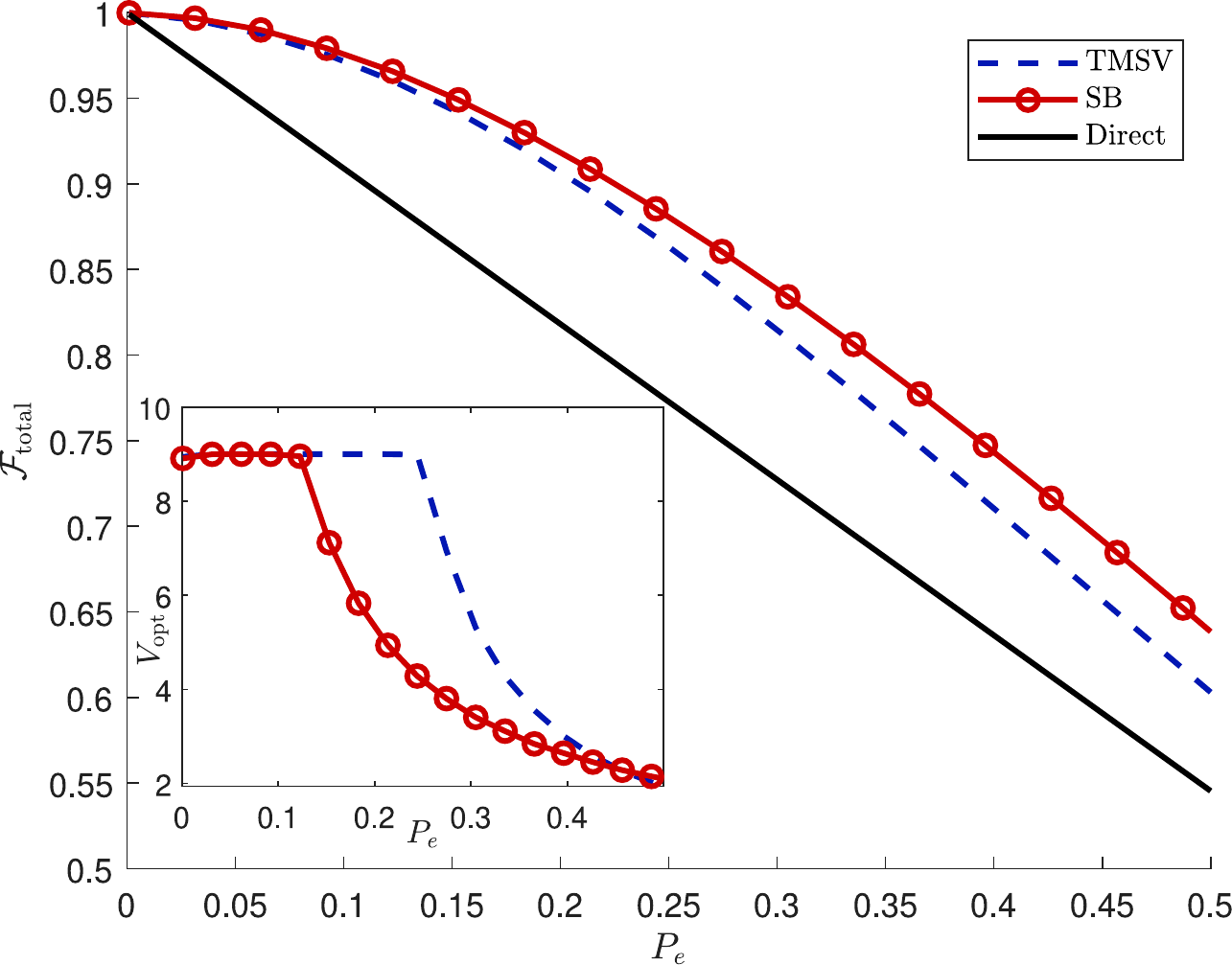}
  \caption{Fidelites obtained for coherent states transmitted through an erasure channel using the error correction code, with TMSV and SB states used as the entangled state. (inset) The optimal value of $V$ used in both entangled states.}
  \label{fig:fid}
\end{figure}

In a realistic scenario, where any number of erasures may affect the encoded state, the optimal value $V_\mathrm{opt}$ must be used to maximize the total fidelity obtained. 
The total fidelity is obtained from the individual fidelities corresponding to each combination of modes erased, weighted by their respective probability of occurrence, as
\begin{align}
  \mathcal{F}_\mathrm{total} &= (1 - P_e)^3 + P_e(1 - P_e)^2 (\mathcal{F}_{1} + \mathcal{F}_{2} + 1) \\ \nonumber
  &~~ + P_e^2(1 - P_e) (\mathcal{F}_{13} + \mathcal{F}_{23} + \mathcal{F}_{\ket{0}}) + P_e^3\mathcal{F}_{\ket{0}}
\end{align}
where $\mathcal{F}_\mathbf{x}$, corresponding to the mean fidelity (Eq.~\ref{eq:avefidelity}) with $\mathbf{x}$ representing the erased modes, and $\mathcal{F}_{\ket{0}}$ corresponds to the mean fidelity between input states and the vacuum.
In Fig.~\ref{fig:fid} the fidelities after the error correction are presented. The results are compared with the fidelities obtained via direct transmission through the channel.
The fidelity is optimized following the procedure discussed in Section~II.A.
We see that the error correction code increases the fidelity of transmitted coherent states for erasure error rates up to 70\%.
Moreover, using the non-Gaussian state improves the fidelities over those acquired using the TMSV state.
Remarkably, once optimized, the non-Gaussian state requires a lower amount of entanglement (lower value of $V$) compared to the TMSV (Fig.~\ref{fig:fid} inset). This could prove advantageous in a scenario where the ability to produce highly-squeezed states is limited.

\subsection{Optimization vs post-selection}
It is also worth considering a simpler protocol that does not involve optimizing the system. A valid protocol would be to post-select the output states to only use those that present only a single erasure, or no erasure at all. In this case, no optimization of the entangled state is required, since the best strategy to recover the signal is to maximize $V$ within device limitations.

\begin{figure}
  \centering
  \includegraphics[width=.48\textwidth]{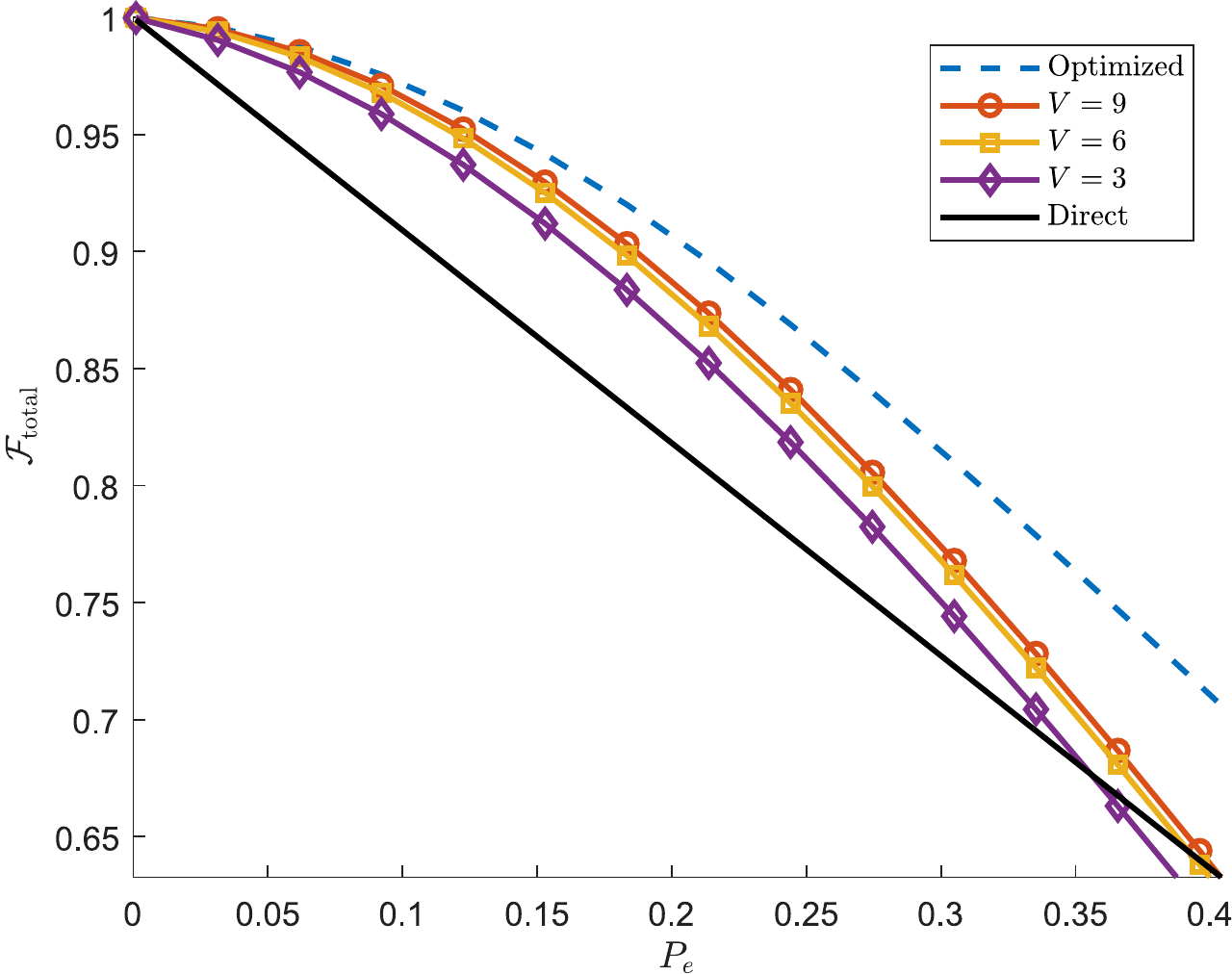}
  \caption{Total fidelities obtained for the post-selection protocol. A TMSV state is used as the entangled state. For comparison, the total fidelities obtained using the optimized protocol with a TMSV state are also presented.}
  \label{fig:opti}
\end{figure}

To perform a fair comparison between the post-selection and the optimization protocol, the mean fidelities are compared. This means that in the post-selection protocol the mean fidelity obtained is weighted by the probability that at most only one erasure occurs, $p_\mathrm{success} {=} (1-P_e)^3 + 3P_e(1-P_e)^2$.
The results are presented in Fig.~\ref{fig:opti}. The fidelities shown in this figure are computed for the post-processing protocol using a TMSV state with different values of $V$.
The fidelities obtained from the post-selection protocol do not exceed the fidelity obtained by the optimized protocol. Nonetheless, we see that for low erasure rates the results of the post-selection protocol are considerably closer to those obtained from the optimized protocol. This means that in a practical implementation of the erasure code the post-selection protocol could still be used effectively if the erasure rate is low.

\section{Conclusions}
 The error correction code studied in this paper represents a practical solution to erasures that could be implemented in future implementations - especially satellite-based implementations in which erasures (and partial erasures) are not uncommon. 
We showed that there existed several free parameters associated with our erasure code that could be optimized.  After such optimization, the fidelities of transmitted coherent states over the erasure channel showed a considerable increase over direct transmission. A simpler protocol was investigated in which the parameters of the code were not optimized, instead, the results were post-selected to consider only when a single (or no) erasure occurs. For low error rates, the post-selected protocol provided similar performance to the optimized protocol while being considerably less complex. 
Our erasure code was further enhanced by considering non-Gaussian states in the encoding process.
Typically, enhancements in the fidelity of $20\%$ were found if non-Gaussian states were used in the encoding process. 

\begingroup
\raggedright

\bibliographystyle{unsrt}
\bibliography{bib}
\endgroup

\end{document}